\documentclass{PoS}
\usepackage{epsfig}
\usepackage{cite}

\title{Exploration of the electric spin polarizability of the neutron in
lattice QCD}

\ShortTitle{Exploration of the electric spin polarizability of the neutron
in lattice QCD}

\author{\speaker{Michael Engelhardt}
\\
        Department of Physics, New Mexico State University, Las Cruces,
        NM 88003, USA\\
        E-mail: \email{engel@nmsu.edu}}

\abstract{A scheme to calculate the electric spin polarizability of the
neutron, based on a four-point function approach to the background field
method, is presented. The connected contributions to this spin polarizability
are evaluated within a mixed action calculation employing domain wall
valence quarks on MILC asqtad sea quark ensembles. Results are reported
for two pion masses, 759 MeV and 357 MeV.}

\FullConference{XXIX International Symposium on Lattice Field Theory \\
                 July 10 - 16, 2011\\
                 Squaw Valley, Lake Tahoe, California}

\begin{document}

\section{Introduction}
Polarizabilities represent fundamental properties of hadrons, encoding
their linear response to externally applied fields. Experimentally, they
manifest themselves, e.g., in the non-Born part of the low-energy Compton
scattering amplitude. While the leading low-energy response is controlled
by the static polarizabilities found in the presence of constant external
fields, at subsequent orders of a derivative expansion, the effective hadron
Hamiltonian becomes sensitive to temporal and spatial structures in the
applied fields. The present study focuses, specifically, on weak, spatially
constant external electric fields $\vec{E} $ applied to a neutron. When the
fields are sufficiently weak, nonlinear effects, i.e., terms in the
effective neutron Hamiltonian of higher order than quadratic in the
external field, can be neglected, and, to zeroth order in spatial
derivatives as well as second order in temporal derivatives of $\vec{E} $,
the neutron mass shift can be expanded as
\begin{equation}
\Delta m = m(\vec{E} )-m(0) = -\frac{1}{2} \alpha_{E} \vec{E}^2 \
-\frac{1}{2} \gamma_{E1} \vec{\sigma } \cdot
(\vec{E} \times \dot{\vec{E} } ) \
-\frac{1}{2} \alpha_{E\nu } \dot{\vec{E} }^2 \ + \ldots
\label{heff}
\end{equation}
The static electric polarizability $\alpha_{E} $ has been investigated
in a number of lattice studies, cf., e.g.,
\cite{wilepap,shintani,elpol,polprog,alexan,detlat}.
Here, the goal is to perform a first evaluation of the electric spin
polarizability $\gamma_{E1} $ within lattice QCD. An effort in this direction
is particularly timely in view of an ongoing experimental program at
HI$\gamma $S to measure this polarizability in isolation for the first
time, albeit in a proton; previously, only the forward and backward spin
polarizabilities $\gamma_{0} $, $\gamma_{\pi } $ of the proton
\cite{spexp1,spexp2}, which contain $\gamma_{E1} $ in combination with
other polarizabilities, had been accessed experimentally.

\section{External field and four-point function scheme}
\begin{figure}[h]
\vspace{-0.5cm}
\hspace{1.1cm} \epsfig{file=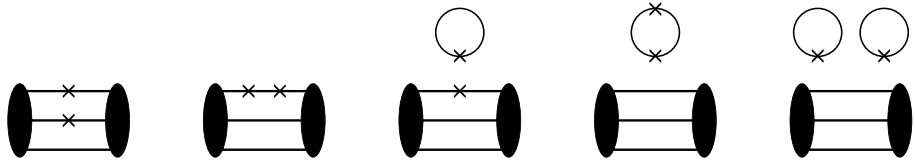,width=13cm}
\vspace{-0.9cm}
\caption{Diagrams contributing to the neutron mass shift; crosses denote
$j\cdot A$ insertions.}
\label{diagrams}
\vspace{-0.15cm}
\end{figure}

Consider a spatially constant external electromagnetic gauge field $A_i $
generating an electric field $E_i = \partial_0 A_i $. It enters the
lattice link variables $U_i $ as an additional phase,
$U_i \rightarrow \exp (iaq A_i ) \cdot U_i $, where $a$ denotes the
lattice spacing and $q$ the quark electric charge matrix. To evaluate only
contributions to the neutron mass shift quadratic in $\vec{E}$, it is
sufficient to expand the relevant neutron two-point function in powers
of $\vec{E}$ from the outset. Thus, one can perform a fully dynamical
calculation using existing unperturbed gauge ensembles:
Expanding $\exp (iaq A_i ) = 1+iaq A_i -(a^2 q^2 /2) A_i^2 +\ldots $ and
inserting into the lattice action decomposes the latter into an
unperturbed part and an external field-dependent part, $S=S_0 + S_{ext} $,
where $S_{ext} $ is essentially a standard $j\cdot A$ coupling of the
external field to the vector current\footnote{Additional
contact terms proportional to $A_i^2 $, cf.~\cite{elpol,polprog},
can be excluded in the present context, as seen below.}.
As a consequence, the neutron two-point function expands as
\begin{equation}
\left\langle N_{\beta } (y) \bar{N}_{\alpha } (x) \right\rangle =
\begin{array}{c}
\int [DU] [D\psi ] [D\bar{\psi } ] \ \exp (-S_0 ) \
\left( 1 - S_{ext} + S_{ext}^{2} /2 + \ldots \right)
N_{\beta } (y) \bar{N}_{\alpha } (x) \\
\hline
\int [DU] [D\psi ] [D\bar{\psi } ] \ \exp
(-S_0 ) \ \left( 1 - S_{ext} + S_{ext}^{2} /2 + \ldots \right)
\end{array}
\end{equation}
Performing the quark integrations
yields the diagrammatic representation for the contributions to the
neutron two-point function quadratic in $\vec{A} $ depicted in
Fig.~\ref{diagrams}.
In the present study, only the connected contributions in Fig.~\ref{diagrams}
were evaluated. The four-point function scheme at which one has thus arrived
is particularly suited to isolate the spin polarizability $\gamma_{E1} $
from the other contributions appearing in (\ref{heff}): As will be specified
in more detail below, each vertex in Fig.~\ref{diagrams} is associated with
an external field insertion $\vec{A} = A_1 \vec{e}_{1} + A_2 \vec{e}_{2} $;
however, in view of (\ref{heff}), only contributions $\sim A_1 A_2 $
\linebreak are of \pagebreak
interest, which can be isolated by populating one vertex with only $A_1 $
and the other vertex with only $A_2 $ in each diagram (and, thus, any
contact terms proportional to $A_i^2 $ can also be disregarded, as noted
above).  This not only discards the signals from the contributions
controlled by $\alpha_{E} $ and $\alpha_{E\nu } $ in (\ref{heff}), but also
their statistical noise\footnote{Note that evaluating the difference
between the mass shifts obtained with neutrons polarized in two
opposing directions does not achieve this; that
eliminates the signal, but not the noise associated with the $\alpha_{E} $
and $\alpha_{E\nu } $ contributions.}. The very specific control over the
different contributions afforded by the present scheme
thus aids in reducing the numerical uncertainties.

Choosing the neutron state to be polarized in the 3-direction,
a suitable external gauge field is
\begin{equation}
\vec{A} = ( a_1^0 + a_1^1 t ,\ a_2^0 + a_2^1 t + a_2^2 t^2 ,\ 0) \ ,
\label{gena}
\end{equation}
which generates a constant $E_1 $ and an $E_2 $ varying linearly with time,
as called for by (\ref{heff}). An ambiguity remains in the choice of the
constants $a_1^0 $, $a_2^0 $ and $a_2^1 $; fixing $E_1 $ and $\dot{E}_2 $
only determines $a_1^1 $ and $a_2^2 $. In particular, the finite spatial
lattice torus is distinct from an infinite spatial domain in that the
constant potentials $a_i^0 $ cannot be wholly eliminated
using gauge transformations. Due to the torus boundary
conditions, only the residual discrete invariance
$A_i \longrightarrow A_i +2\pi /q_{min} L$ remains, where $L$ is the
spatial extent of the torus and $q_{min}$ the smallest unit of electrical
charge in the theory. Fractionally different $a_i^0 $ correspond to
different physics; namely, they represent flavor-dependent Bloch momenta
in view of the minimal substitution $p_i \rightarrow (p_i -q a_i^0 )$.
Thus, despite the dependences on the constants $a_1^0 $, $a_2^0 $
and $a_2^1 $ being partially related to one another via space-time symmetries,
the neutron mass shift is influenced by additional physics distinct from
the spin polarizability effect,
\begin{equation}
\Delta m = c_1 \vec{\sigma } \cdot (\vec{A} \times \vec{E} ) \
-\frac{1}{2} \gamma_{E1} \vec{\sigma } \cdot
(\vec{E} \times \dot{\vec{E} } )
\label{haeff}
\end{equation}
even after having restricted the calculation to contributions $\sim A_1 A_2 $.
The additional Bloch momentum dependence\footnote{\label{foot3}Note that a
separate dependence of the form $\vec{A} \times \dot{\vec{E} } $ can be
excluded as follows: Working in Euclidean space, each time derivative acquires
a factor $i$ compared to Minkowski space. The terms in (\ref{haeff}) both
contain an odd number of time derivatives, and $\Delta m $ will
correspondingly be extracted from the imaginary part of the neutron two-point
function. By contrast, any putative physical $\vec{A} \times \dot{\vec{E} } $
dependence cannot (and will explicitly be seen not to) arise in the imaginary
part.} must be disentangled from the polarizability effects in order to
extract $\gamma_{E1} $. To this end, several choices of external
fields of the type (\ref{gena}) will be considered and contrasted below.

\section{Extracting the neutron mass shift in the adiabatic approximation}
In the presence of external fields of the form (\ref{gena}), the Hamiltonian
of the system under consideration is time-dependent. One facet of this
which deserves to be noted is the connection to the Bloch momenta discussed
above. One can, e.g., write $A_1 $ in (\ref{gena}) in two equivalent ways:
$A_1 = a_1^0 + a_1^1 t \equiv a_1^1 (t-t_0 )$, i.e., a translation of the
system by the time $t_0 $ corresponds to the introduction of a Bloch momentum
$a_1^0 = -a_1^1 t_0 $. This, of course, simply expresses the fact that the
electric field $E_1 $ accelerates or decelerates the Bloch currents.
Correspondingly, the Bloch momentum dependence of the neutron mass shift
\pagebreak
will manifest itself in a characteristic time dependence of the neutron
two-point function, as will be seen below.

In general, the time evolution operator for a time-dependent Hamiltonian
acquires an intricate time dependence. Here, an adiabatic approximation
will be adopted to interpret the measured behavior of the neutron two-point
function: Since the external field can be taken to be arbitrarily weak, it
seems plausible to consider the case that the strong dynamics equilibrate
the system rapidly on the scale of the temporal variation of the external
field. In this situation, the behavior of the two-point function (now
projected onto zero momentum states polarized in the 3-direction) is, at
times $t$ sufficiently large for excited states to have decayed,
\begin{equation}
G_{\uparrow} (p=0,t) =
\sum_{\vec{y} } \ \mbox{Tr}
\left( \frac{1+\gamma_{4} }{2} (1-i\gamma_{3} \gamma_{5} )
\left\langle N (y) \bar{N} (x) \right\rangle \right)
\ \ = \ \
W \ \exp \left( -\int dt^{\prime } \, m(t^{\prime } ) \right)
\end{equation}
Inserting expansions of the normalization and the mass\footnote{Note
that the omission of a linear term $m^{(1)} $ in the ansatz
(\ref{massexp}) implies that the absence of a permanent net electric current
or electric dipole moment in an unperturbed neutron has already been put in.}
in the external fields (where the superscript specifies the order in the
external field),
\begin{eqnarray}
W &=& W_0 + W^{(1)} [\vec{A} (t)] + W^{(2)} [\vec{A} (t)] +\ldots \\
m &=& m_0 + m^{(2)} [\vec{A} (t)] +\ldots
\label{massexp}
\end{eqnarray}
one obtains specifically for the contribution quadratic in the external
field
\begin{equation}
G^{(2)}_{\uparrow} (p=0,t) \ \ = \ \
W_0 \ \exp (-m_0 \, t) \left( \frac{W^{(2)} [\vec{A} (t)]}{W_0 }
-\int dt^{\prime } \, m^{(2)} [\vec{A} (t^{\prime } )] \right)
\label{corr2}
\end{equation}
In view of this, the mass shift $m^{(2)} [\vec{A} (t)]$ can be obtained,
after dividing out the unperturbed correlator
$G^{(0)} (p=0,t) = W_0 \ \exp (-m_0 \, t)$,
from the temporal slope of the ratio {\em only at a stationary point} of
the time evolution, where the time dependences of $W^{(2)} [\vec{A} (t)]$
and $m^{(2)} [\vec{A} (t)]$ are relegated $\mbox{\hspace{1.5cm} } $
\vspace{-0.2cm}
\begin{table}[h]
\hspace{0.05cm}
\begin{tabular}{|c|c|c|c|}
\hline
$am_l $ & $am_s $ & $ m_{\pi } $ & \# configs \\
\hline \hline
0.01 & 0.05 & 357 MeV & 448 \\
\hline
0.05 & 0.05 & 759 MeV & 425 \\
\hline
\end{tabular}
\caption{$N_f =2+1$, $20^3 \times 64 $ MILC asqtad
\hspace{8.6cm} $\mbox{\ \ } $
ensembles with $a=0.124\, \mbox{fm} $ used in the
\hspace{8.43cm} $\mbox{\ \ } $
present investigation.}
\label{tabmilc}
\end{table}
\vspace{-4cm}

\hspace{6.3cm} \parbox{8.02cm}{to quadratic order in $t$.
Physically, this occurs when the quark Bloch currents cease to flow and turn
around into the opposite direction due to the forcing by the external
electric field. Only at such a stationary point do the strong dynamics have
the opportunity to form a bona fide neutron, the polarizability of which
can thus be extracted only in the vicinity \linebreak of that particular
point in time.}

\section{Numerical results}
\begin{figure}[t]
\vspace{-0.1cm}
\epsfig{file=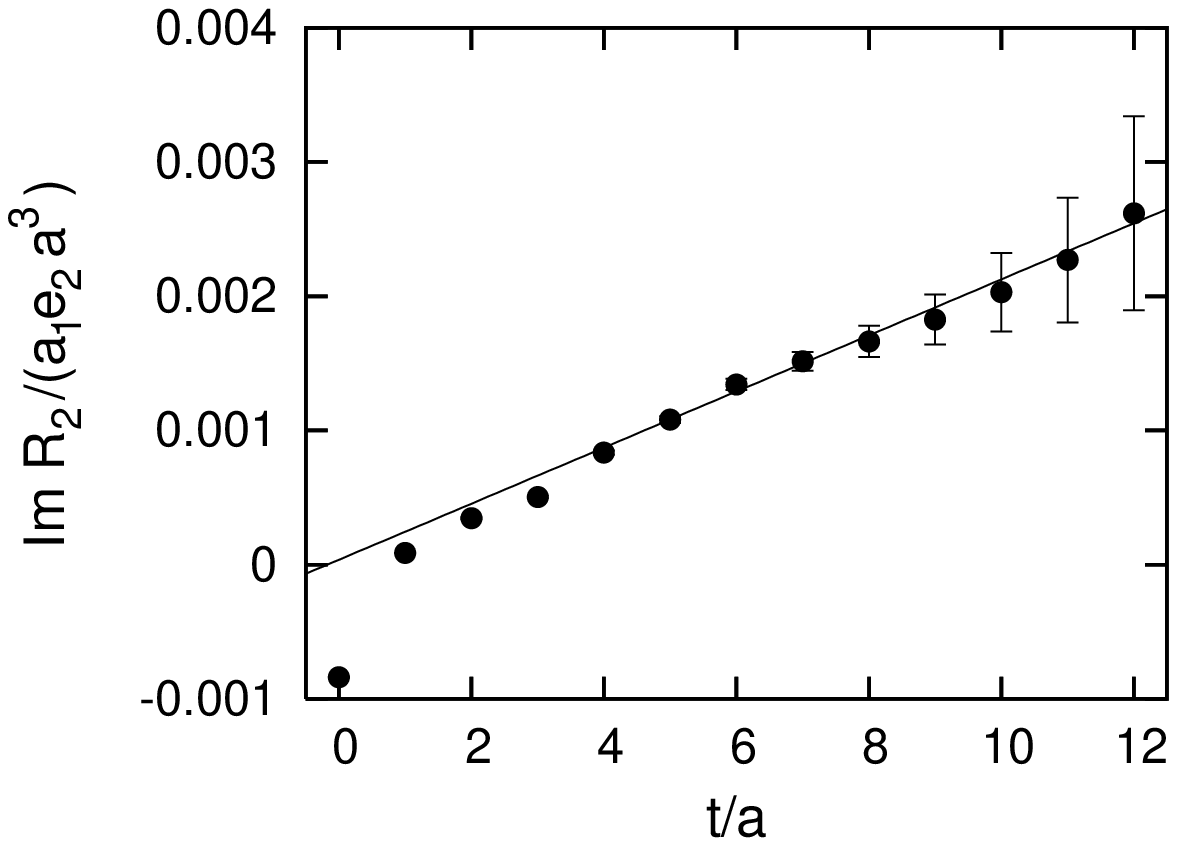,width=4.9cm,angle=-90} \hspace{0.1cm}
\epsfig{file=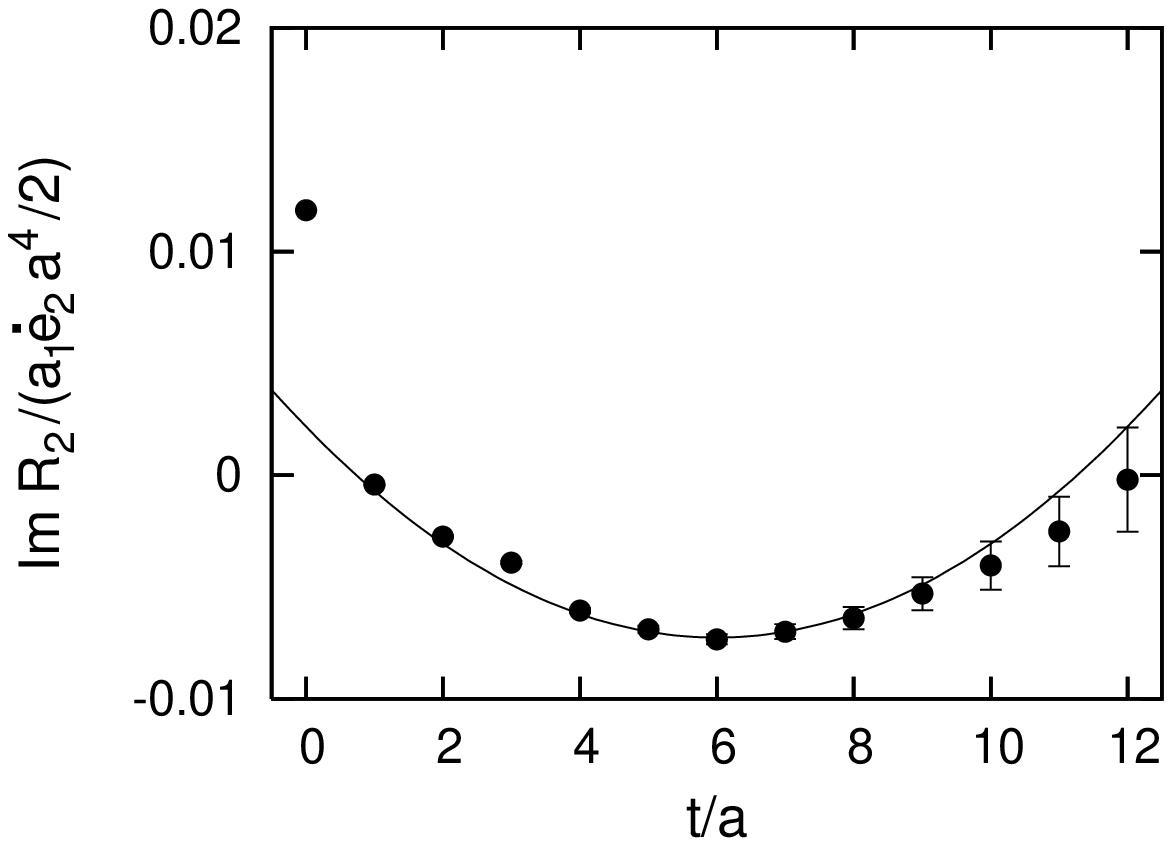,width=4.9cm,angle=-90}
\caption{Neutron two-point function ratio $R_2 $ obtained in the
$m_{\pi } =759\, \mbox{MeV} $ ensemble, in the presence of the external
fields $\vec{A} = (a_1 ,\ e_2 (t-t_0 ) ,\ 0)$ (left) and
$\vec{A} = (a_1 ,\ \dot{e}_2 (t-t_0 )^2 /2 ,\ 0)$ (right), where
$t_0 =6a$. The neutron source is located at $t=0$, electromagnetic
fields are in Gaussian units. Linear and quadratic fits, respectively,
were performed using the fit range $t/a \in [4,8]$.}
\label{fig759ab}
\end{figure}

\begin{figure}[h]
\vspace{0.15cm}
\hspace{0.05cm}
\epsfig{file=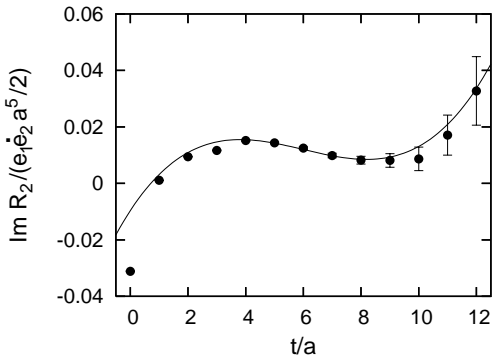,width=4.9cm,angle=-90}
\caption{Neutron two-point function ratio $R_2 $ ob-
\hspace{7.53cm} $\mbox{\ \ } $
tained in the $m_{\pi } =759\, \mbox{MeV} $ ensemble, given the ex-
\hspace{7.62cm} $\mbox{\ \ } $
ternal field $\vec{A} = (e_1 (t-t_0 ) ,\ \dot{e}_2 (t-t_0 )^2 /2 ,\ 0)$,
where
\hspace{7.58cm} $\mbox{\ \ } $
$t_0 =6a$. The neutron source is located at $t=0$, and
\hspace{7.58cm} $\mbox{\ \ } $
electromagnetic fields are in Gaussian units. Cubic
\hspace{7.54cm} $\mbox{\ \ } $
fit was performed using the fit range $t/a \in [4,8]$.}
\label{fig759c}
\vspace{-8.67cm}
\end{figure}

\begin{figure}[t]
\vspace{-0.2cm}
\epsfig{file=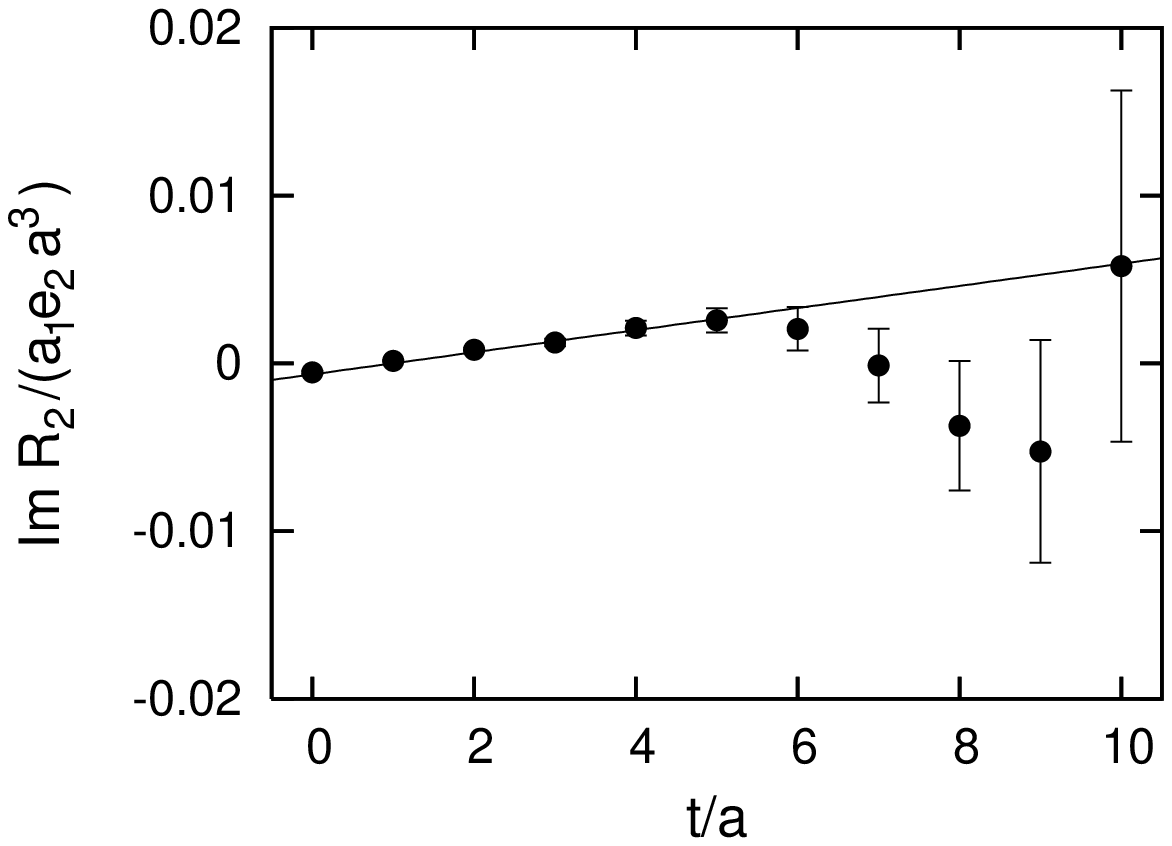,width=4.9cm,angle=-90} \hspace{0.1cm}
\epsfig{file=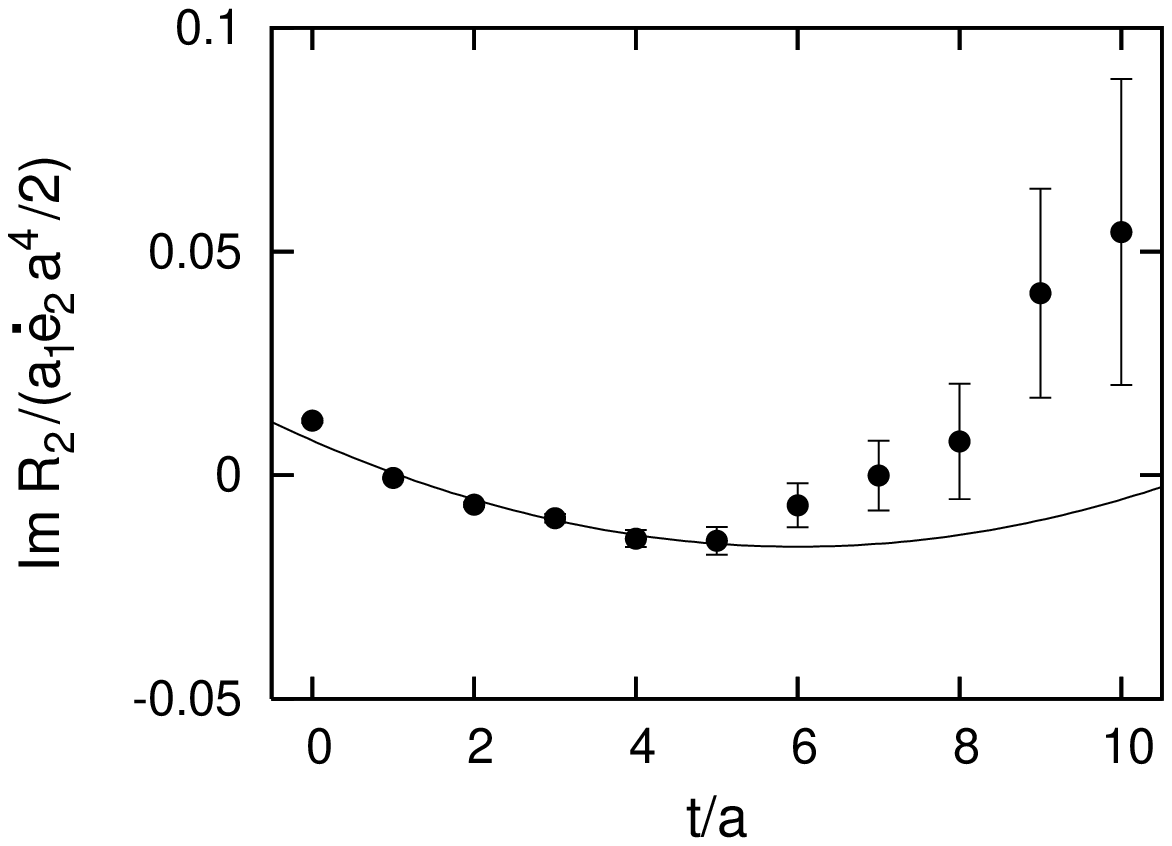,width=4.9cm,angle=-90}
\caption{Correlator ratios $R_2 $ analogous to Fig.~2,
except in the $m_{\pi } =357\, \mbox{MeV} $ ensemble, with fits
performed using the fit range $t/a \in [3,5]$.}
\label{fig357ab}
\vspace{-0.1cm}
\end{figure}

\begin{figure}[h]
\vspace{-0.54cm}
\hspace{0.05cm}
\epsfig{file=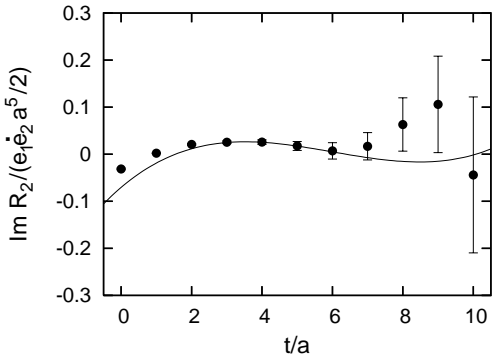,width=4.9cm,angle=-90}
\caption{Correlator ratio $R_2 $ analogous to Fig.~3,
\hspace{7.78cm} $\mbox{\ \ } $
except\, in\, the\, $m_{\pi } =357\, \mbox{MeV} $\, ensemble,\, with\, fit
\hspace{7.8cm} $\mbox{\ \ } $
performed using the fit range $t/a \in [3,5]$.}
\label{fig357c}
\vspace{-6.94cm}
\end{figure}

The neutron two-point function was evaluated using the MILC ensembles listed
in Table~\ref{tabmilc}.
Each configuration was HYP-smeared and chopped into two $20^3 \times 32 $
sublattices with temporal Dirichlet boundary conditions. Domain wall valence
quarks were employed. To improve statistics, averages over neutron spin in
the positive and negative 3-directions were taken, as well as over the
external field (\ref{gena}) and its rotation by $\pi /4$ around the
3-axis. Figs.~\ref{fig759ab}-\ref{fig357c} display results for the ratio
$R_2 (t) = G_{\uparrow }^{(2)} (p=0,t) / G^{(0)} (p=0,t)$ for several cases
of external field.

Consider first Figs.~\ref{fig759ab} and \ref{fig759c}, which pertain to the
case of pion mass $m_{\pi } = 759$ MeV. To exhibit the different effects at
play, Fig.~\ref{fig759ab}~(left) shows the result of merely using an
\pagebreak
external field of the form $\vec{A} = (a_1 ,\ e_2 (t-t_0 ) ,\ 0)$, which is
expected to yield a time-independent effective dynamics isolating
the $\vec{\sigma } \cdot (\vec{A} \times \vec{E} )$ dependence in
(\ref{haeff}). Indeed, the correlator ratio $R_2 (t)$ exhibits linear
behavior, corresponding to a constant neutron mass
shift determined by the slope of $R_2 (t)$, cf.~(\ref{corr2}).
Continuing with the more complex external field
$\vec{A} = (a_1 ,\ \dot{e}_2 (t-t_0 )^2 /2 ,\ 0)$,
the expected behavior of the correlator ratio $R_2 (t)$ is parabolic
in the vicinity of the stationary point $t=t_0 $, since now the electric
field entering the $\vec{\sigma } \cdot (\vec{A} \times \vec{E} )$
dependence of (\ref{haeff}) is linear in time, crossing zero at $t=t_0 $.
On the other hand, an additional mass shift proportional to
$\vec{\sigma } \cdot (\vec{A} \times \dot{\vec{E} } )$ would manifest
itself as an additional linear term in $R_2 (t)$, with the effect of
shifting the parabola away from $t=t_0 $. Fig.~\ref{fig759ab}~(right),
depicting the corresponding numerical result, corroborates the expected
behavior; there is no shift of the parabolic time dependence away from
$t=t_0 $, and therefore no additional dependence on the combination
$\vec{\sigma } \cdot (\vec{A} \times \dot{\vec{E} } )$, as expected on
general grounds, cf.~footnote~\ref{foot3} further above. One can moreover
verify that the quadratic coefficient of the parabola in
Fig.~\ref{fig759ab}~(right) is compatible with the slope in
Fig.~\ref{fig759ab}~(left) within numerical error, consistent with
the above interpretation.

\hspace{7.3cm} \parbox{7.02cm}{Finally, employing the full external field
$\vec{A} = (e_1 (t-t_0 ) ,\ \dot{e}_2 (t-t_0 )^2 /2 ,\ 0)$,
one would expect to see cubic behavior of $R_2 (t)$ around the stationary
point $t=t_0 $, since now the time dependences of $A$ and
$E$ in $\vec{\sigma } \cdot (\vec{A} \times \vec{E} )$
combine to render the latter proportional to $(t-t_0 )^2 $.
If this were the only contribution to the neutron mass shift, the slope
of $R_2 (t)$ at the inflection point $t=t_0 $ would vanish; on the other
hand, an additional contribution to the mass shift proportional to
$\vec{\sigma } \cdot (\vec{E} \times \dot{\vec{E} } )$ would manifest
itself as an additional linear term in $R_2 (t)$, the slope of which
could therefore be read off at the inflection point of the overall cubic
behavior. This slope, of course, determines the electric \hspace{-0.039cm}
spin \hspace{-0.039cm} polarizability \hspace{-0.039cm} $\gamma_{E1} $
\hspace{-0.039cm} of \hspace{-0.039cm} the \hspace{-0.039cm} neutron}
which is the ultimate objective of the analysis. Indeed, the corresponding
numerical result for $R_2 (t)$ shown in Fig.~\ref{fig759c} exhibits the
expected behavior, including a nonvanishing slope at the stationary point
$t=t_0 $. From this slope, one extracts the value
$\gamma_{E1} = 0.0057(8) \cdot 10^{-4} \, \mbox{fm}^{4} $, from connected
contributions only, at the pion mass $m_{\pi } = 759$ MeV.

Turning to Figs.~\ref{fig357ab} and \ref{fig357c}, which present analogous
results for the lighter pion mass $m_{\pi } = 357$ MeV, the principal
difference lies in the substantially larger statistical uncertainties, which
preclude a clear identification of the functional form of the neutron
correlator beyond the time $t=5a$ in the plots. As a result, fits in a
symmetric time range around the stationary point $t=6a$, as employed in
the $m_{\pi } = 759$ MeV case, are not viable at the present level of
statistics. Instead, pending improvement of the numerical accuracy, a fit
in the time range $t/a\in [3,5]$ was used to identify the preliminary value
$\gamma_{E1} = 0.031(21) \cdot 10^{-4} \, \mbox{fm}^{4} $, from connected
contributions only, at the pion mass $m_{\pi } = 357$ MeV. Nevertheless,
the data depicted in Figs.~\ref{fig357ab} and \ref{fig357c} display behavior
entirely analogous to the one seen in the heavier pion mass case for small
times, where the numerical fluctuations are under control.

\begin{figure}[h]
\vspace{-0.18cm}
\hspace{0.05cm}
\epsfig{file=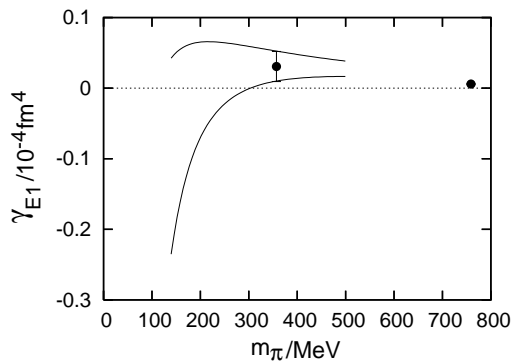,width=4.9cm,angle=-90}
\caption{Measured lattice data in relation to par-
\hspace{7.7cm} $\mbox{\ \ } $
tially quenched chiral perturbation theory, cf.~text.}
\label{figeft}
\vspace{-6.67cm}
\end{figure}

Finally, Fig.~\ref{figeft} relates the two data points for the spin
polarizability $\gamma_{E1} $ extracted above to partially
quenched chiral perturbation theory \cite{detcsb}, adjusted to match the
present connected diagram calculation. The corresponding expression
depends on a number of low energy constants, all but one of which are
comparatively well determined from either experiment or previous lattice
calcula- $\mbox{\hspace{0.5cm}}$

\hspace{7.3cm} \parbox{7.02cm}{tions on the same ensembles. The lone poorly
determined parameter, $g_1 $, has been adjusted such as to generate the two
curves in Fig.~\ref{figeft} which pass through the upper and lower ends of
the error bar of the $\gamma_{E1} $ measurement at $m_{\pi } = 357$ MeV. This
yields a rather stringent estimate of
$g_1 = -0.15\,\,^{+0.01+0.04}_{-0.01-0.10}\,$,
where the first uncertainty quantifies the spread depicted in
Fig.~\ref{figeft}, and the second one was obtained by varying the other
low energy constants within reasonable bounds. Note that errors due to
the truncation of the chiral expansion were not estimated.}

\newpage

\section{Summary}
The present investigation provides first lattice QCD results for the
electric spin polarizability $\gamma_{E1} $ of the neutron, albeit
including connected diagrams only. The extraction of this quantity was
facilitated by the adopted four-point function approach, which allows
for the exact elimination of other contributions to the neutron mass
shift in the external electric field considered, namely, ones associated
with the static polarizability $\alpha_{E} $ and the dispersion
polarizability $\alpha_{E\nu} $. In addition, the effects of constant
external gauge fields, corresponding to quark Bloch momenta on a finite
spatial torus, had to be carefully disentangled from the spin
polarizability effect itself. A clear signal for $\mbox{\hspace{0.5cm}}$

\hspace{7.3cm} \parbox{7.02cm}{
the connected contribution to $\gamma_{E1} $ was obtained at the pion mass
$m_{\pi } = 759\, \mbox{MeV} $; also a preliminary extraction at
$m_{\pi } = 357\, \mbox{MeV} $ proved possible, although a refinement of
the statistical accuracy is desirable. The most striking feature of the
connected contribution to $\gamma_{E1} $ obtained here is its smallness
compared to the result expected for full QCD from chiral perturbation
theory \cite{detcsb}, namely, a negative $\gamma_{E1} $ larger in modulus
by roughly two orders of magnitude. This suggests that the electric spin
polarizability $\gamma_{E1} $ of the neutron is dominated by the
disconnected contributions.}

\vspace{0.099cm}

\section*{Acknowledgments}
Discussions with F.~X.~Lee, presenting related work at this conference,
and W.~Detmold are gratefully acknowledged. Computations were performed using
Chroma \cite{chroma} on U.S.~DOE/USQCD resources at Fermilab.
This work was supported by U.S.~DOE grant DE-FG02-96ER40965.

\end{document}